\documentclass[conference]{IEEEtran}
\IEEEoverridecommandlockouts
\usepackage{cite}
\usepackage{amsmath,amssymb,amsfonts}
\usepackage{algorithmic}
\usepackage{graphicx}
\usepackage{textcomp}
\usepackage{xcolor}
\usepackage{hyperref}
\usepackage{braket}

\usepackage{subcaption} 
\def\BibTeX{{\rm B\kern-.05em{\sc i\kern-.025em b}\kern-.08em
    T\kern-.1667em\lower.7ex\hbox{E}\kern-.125emX}}
\begin{document}

\title{Can Classical Initialization Help Variational Quantum Circuits Escape the Barren Plateau? \thanks{The views expressed in this article are those of the authors and do not represent the views of Wells Fargo. This article is for informational purposes only. Nothing contained in this article should be construed as investment advice. Wells Fargo makes no express or implied warranties and expressly disclaims all legal, tax, and accounting implications related to this article.}}
\author{\IEEEauthorblockN{1\textsuperscript{st} Yifeng Peng}
\IEEEauthorblockA{\textit{School of Engineering and Science} \\
\textit{Stevens Institute of Technology}\\
Hoboken, USA \\
ypeng21@stevens.edu}
\and
\IEEEauthorblockN{2\textsuperscript{nd} Xinyi Li}
\IEEEauthorblockA{\textit{School of Engineering and Science} \\
\textit{Stevens Institute of Technology}\\
Hoboken, USA \\
xli215@stevens.edu}
\and
\IEEEauthorblockN{3\textsuperscript{rd} Zhemin Zhang}
\IEEEauthorblockA{\textit{ECSE Department} \\
\textit{Rensselaer Polytechnic Institute}\\
Troy, USA \\
zhangz29@rpi.edu} 
\and
\IEEEauthorblockN{4\textsuperscript{th} Samuel Yen-Chi Chen}
\IEEEauthorblockA{\textit{Wells Fargo} \\
New York, USA \\
ycchen1989@ieee.org}
\and
\IEEEauthorblockN{5\textsuperscript{th} Zhiding Liang}
\IEEEauthorblockA{\textit{Computer Science Department} \\
\textit{Rensselaer Polytechnic Institute}\\
Troy, USA \\
liangz9@rpi.edu}
\and
\IEEEauthorblockN{6\textsuperscript{th} Ying Wang}
\IEEEauthorblockA{\textit{School of Engineering and Science} \\
\textit{Stevens Institute of Technology}\\
Hoboken, USA \\
ywang6@stevens.edu}
}
\maketitle
\begin{abstract}
Variational quantum algorithms (VQAs) have emerged as a leading paradigm in near-term quantum computing, yet their performance can be hindered by the so-called barren plateau problem, where gradients vanish exponentially with system size or circuit depth. While most existing VQA research employs simple Gaussian or zero-initialization schemes, classical deep learning has long benefited from sophisticated weight initialization strategies such as Xavier, He, and orthogonal initialization to improve gradient flow and expedite convergence. In this work, we systematically investigate whether these classical methods can mitigate barren plateaus in quantum circuits. We first review each initialization's theoretical grounding and outline how to adapt the notions from neural networks to VQAs. We then conduct extensive numerical experiments on various circuit architectures and optimization tasks. Our findings indicate that while the initial heuristics, inspired by classical initialization, yield moderate improvements in certain experiments, their overall benefits remain marginal. By outlining a preliminary exploration plan in this paper, we aim to offer the research community a broader perspective and accessible demonstrations. Furthermore, we propose future research directions that may be further refined by leveraging the insights gained from this work.
\end{abstract}


\section{Introduction}\label{sec:introduction}

Variational quantum algorithms (VQAs) have emerged as a leading framework for leveraging near-term quantum devices to tackle a range of problems in chemistry, optimization, and machine learning \cite{peng2024hyq2, peng2024qrng, peng2024qsco, peng2024quantum,peng2025quantum, peng2024hybrid}. By combining classical optimization methods with parameterized quantum circuits (PQCs), VQAs aim to iteratively refine a cost function, often given by the expectation value of a Hamiltonian or an objective operator. While promising in principle, VQAs suffer from the \emph{Barren Plateau} (BP) phenomenon, wherein gradients vanish exponentially as system size grows, severely limiting scalability and performance~\cite{mcclean2018barren,cerezo2021cost,pesah2021absence, larocca2024review}.

In the classical deep learning community, network initialization has long been recognized as critical in addressing vanishing or exploding gradients. He \emph{et al.}~\cite{he2015delving}, LeCun \emph{et al.}~\cite{lecun2002efficient}, Saxe \emph{et al.}~\cite{saxe2013exact} and Glorot and Bengio~\cite{glorot2010understanding} established initialization schemes---collectively known as He, LeCun, Orthogonal and Xavier initializations, respectively---that balance the variance of forward and backward signals, facilitating stable training in multi-layer neural networks. These methods exploit statistical insights about signal propagation to ensure that neither gradients nor outputs become pathologically large or small as depth increases.

\begin{figure}
\centering
\includegraphics[width=0.8\columnwidth]{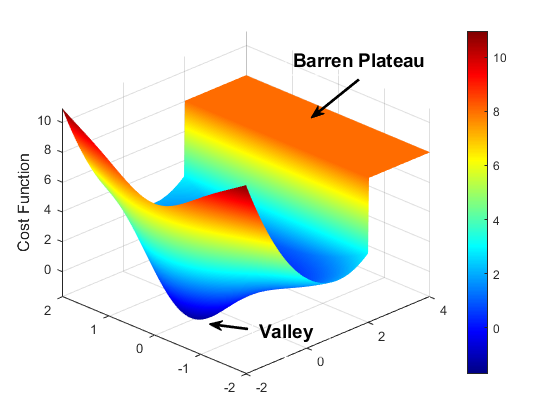} 
\caption{Barren Plateau landscape visualization.}
\label{BPlandscape}
\vspace{-5mm}
\end{figure}

Inspired by the successes of these \emph{classical} initialization methods, a natural question arises: \emph{Can classical initialization strategies help VQCs escape or mitigate Barren Plateaus?} The intuition is that, if suitable parameter distributions can limit the exploration of large portions of the Hilbert space or the concentration of gradients near zero, the exponential decay of gradient norms might be tempered in quantum settings as well. Early results in quantum initialization, such as small-angle parameter distributions~\cite{grant2019initialization} and informed warm-starting~\cite{mari2020transfer, volkoff2021large}, have shown promise,  yet they have not been tested and validated across the same breadth of real-world scenarios that have made He, LeCun, and Xavier methods so foundational in classical practice. This paper proposes initialization methods for quantum circuits inspired by classical approaches. The main contributions of this paper are as follows:
\begin{enumerate}
\item \textbf{Comprehensive Survey of Classical Initialization Schemes:} We provide an overview of widely adopted initialization strategies in classical deep learning (Xavier, He, LeCun, Orthogonal, etc.) and propose heuristic quantum initializations to the VQA.

\item \textbf{Systematic Empirical Evaluation:}
We conduct extensive numerical experiments to compare these classical initialization methods against common Gaussian or zero-initializations in VQA across optimization tasks. 

\item \textbf{Open Directions for Future Research:} We highlight open challenges, such as exploring more nuanced definitions of fan-in/fan-out. By providing a preliminary exploration of classically inspired initialization schemes, we aim to inspire more and better ideas and research in the quantum community.
\end{enumerate}

\section{Background and Related Work}\label{sec:related_work}

This section surveys prior work on BP mitigation with a particular focus on initialization strategies, as well as related insights from circuit design.


\paragraph{Small-Angle and Data-Informed Initialization} Since randomly initialized deep circuits tend to produce states in large, high-entropy regions of Hilbert space~\cite{mcclean2018barren}, \emph{small-angle initializations} have been proposed to keep the circuit near identity-like transformations~\cite{grant2019initialization}. By restricting parameter values to a narrow range, one avoids the rapid spread of amplitudes that can lead to exponentially small gradients. In addition to such random-but-constrained strategies, \emph{data-informed} initialization has also been investigated, where partial classical solutions or known symmetries guide the starting point~\cite{cerezo2021cost}. \emph{Gaussian initializations} have likewise been explored by Zhang \emph{et al.}~\cite{zhang2022escaping}, where parameters are sampled from a carefully chosen Gaussian distribution to systematically reduce barren plateau effects in deeper circuits showing better performance than zero and uniform initialization methods. Collectively, these approaches share the goal of reducing over-randomization at the outset, thereby placing the circuit closer to the target manifold.

\paragraph{Layerwise ``Freezing-Unfreezing'' and Sequential Extensions}
A complementary initialization tactic involves \emph{layerwise training}, wherein the circuit is grown incrementally, and parameters in newly added layers are initialized (and trained) while previously trained layers are ``frozen''~\cite{volkoff2021large,skolik2021layerwise}. This sequential approach prevents the entire circuit from randomizing at once, often retaining gradient signal in earlier layers as deeper layers are introduced. While effective in certain settings (e.g., QAOA and VQE), one must ensure that freezing does not lock in suboptimal parameter values that become difficult to correct in later stages.

\paragraph{Warm-Start and Transfer Learning}
Finally, \emph{warm-start} and \emph{transfer learning} methods—whereby parameters learned on smaller or related problems are adapted to new tasks—can also mitigate BPs by avoiding severe random initialization~\cite{mari2020transfer,cerezo2021cost}. However, the effectiveness of these approaches depends heavily on the similarity between the pretraining and target tasks. Large task discrepancies can lead to suboptimal parameter basins, or even introduce new local minima~\cite{henderson2020quanvolutional,rivera2021avoiding}.

\paragraph{Shallow Circuits and Architectural Constraints}
A common BP-mitigation strategy is to reduce circuit depth and use local measurements \cite{cerezo2021cost}, as shallow architectures—such as QCNNs \cite{pesah2021absence}—limit global entanglement and slow gradient decay. However, shallower ans\"atze often suffer \emph{reachability deficits} \cite{mcclean2018barren,akshay2020reachability}, failing to approximate more complex states.

\paragraph{Symmetry-Preserving and Specialized Ans\"atze}
Embedding known symmetries can further stabilize gradients by restricting the accessible subspace \cite{larocca2022diagnosing,ragone2024lie}, though overly narrow symmetry assumptions risk excluding the true solution manifold \cite{rivera2021avoiding}.

\section{Initialization Methods: Classical Insights, Quantum Heuristics, and Future Optimizations}
\label{sec:init_methods}

In this section, we focus on four widely studied initialization techniques Xavier (Glorot), He, LeCun, and Orthogonal, and discuss their \emph{classical formulations} along with how we \emph{implement} them in this paper as heuristic quantum initializations (Section~\ref{subsec:quantum_impl}). We also outline \emph{potential future enhancements} to better align these methods with quantum scenarios.

\subsection{Xavier (Glorot) Initialization}
\subsubsection{Classical Approach}
Proposed by Glorot and Bengio \cite{glorot2010understanding}, Xavier (or Glorot) initialization aims to keep the variance of layer outputs stable in both forward and backward passes of a neural network. For a classical layer with \(\text{fan\_in}\) inputs and \(\text{fan\_out}\) outputs, we have:
\begin{equation}
    \text{Xavier Normal:}\quad
    w_i \sim \mathcal{N}\!\Bigl(0,\; \tfrac{2}{\text{fan\_in} + \text{fan\_out}} \Bigr),
\end{equation}
\begin{equation}
    \text{Xavier Uniform:}\quad
    w_i \sim \mathcal{U}\!\Bigl(-\,\alpha,\;\alpha\Bigr),
    \;\;
    \alpha = \sqrt{\tfrac{6}{\text{fan\_in} + \text{fan\_out}}},
\end{equation}
where $w_i$ denotes the $i$-th weight parameter, $\mathcal{N}$ refers to the normal distribution, and $\mathcal{U}$ refers to the uniform distribution.

\subsubsection{Proposed Heuristic Quantum Initialization}
\label{subsec:quantum_impl}
We provide two main strategies:

\textbf{Chunk-based Layerwise Approach.} 
We partition the total parameter vector of size \(\,n_{\mathrm{params}}\) into \(\,n_{\mathrm{layer}}\) chunks. Each chunk is treated as if \(\text{fan\_in} = \text{fan\_out} = n\) (the number of qubits), leading to 
\begin{equation}
      \sigma_\ell \;=\; \sqrt{\frac{2}{n + n}}\, 
      \;=\;\sqrt{\frac{1}{n}}\, .
    \end{equation}
We draw random values from \(\mathcal{N}(0,\,\sigma_\ell^2)\) (or the uniform variant) to fill each chunk. After generating all chunks, we concatenate them into the initial parameter vector. This mimics the idea of preserving variance at each (quantum) layer, though we heuristically set \(\text{fan\_in} = \text{fan\_out} = n\).

\textbf{Global Approach (Fan-in = \(\,n_{\mathrm{params}}\)).}
Alternatively, one may ignore layerwise splitting and directly set \(\text{fan\_in} = n_{\mathrm{params}},\, \text{fan\_out} = 1\). Our paper implements simpler variants with no chunk partition, so that
\begin{equation}
    w_i \sim \mathcal{N}\!\Bigl(0,\; \gamma^2 \cdot \tfrac{2}{n_{\mathrm{params}}}\Bigr),
\end{equation}
\begin{equation}
w_i \sim \mathcal{U}\!\Bigl(-\,\gamma\sqrt{\tfrac{6}{n_{\mathrm{params}}}},\;
 \gamma\sqrt{\tfrac{6}{n_{\mathrm{params}}}}\Bigr),
 \end{equation}
where \(\gamma\) is a scale factor discussed in Sec. \ref{subsec:heisenberg}.

\subsubsection{Current Limitations}

\textbf{Simplified \text{fan-in}/\text{fan-out} Definitions:} 
 In classical neural networks, \text{fan\_in} and \text{fan\_out} correspond to well-defined dimensions of incoming/outgoing connections. In VQCs, parameter layers and entangling blocks do not necessarily follow a clear feed-forward structure. Thus, using a single measure (e.g., \text{fan\_in} $= qubits$ and \text{fan\_out} $= n_{params}$) may not capture the nuanced gate connectivity or measurement structure.


\subsubsection{Future Optimizations}
 \textbf{Layer-Specific Refinements.} 
Instead of treating all layers identically, one could adapt \(\text{fan\_in}\) and \(\text{fan\_out}\) based on how many qubits each layer acts upon, especially in circuits with partial entangling gates.

\textbf{Circuit-Depth and Entanglement Structure.} 
In deeper circuits, the effective variance might need further adjustments to avoid barren plateaus \cite{mcclean2018barren} or to account for gate connectivity.
\subsection{LeCun Initialization}
\subsubsection{Classical Approach}
LeCun initialization \cite{lecun2002efficient} is designed for sigmoid or tanh networks, choosing:
\begin{equation}
    \text{LeCun Normal:}\quad
    w_i \sim \mathcal{N}\!\Bigl(0,\; \tfrac{1}{\text{fan\_in}}\Bigr),
\end{equation}
\begin{equation}
    \text{LeCun Uniform:}\quad
    w_i \sim \mathcal{U}\!\Bigl(-\,\delta,\;\delta\Bigr),
    \;\;\delta = \sqrt{\tfrac{3}{\text{fan\_in}}}.
\end{equation}

\subsubsection{Proposed Heuristic Quantum Initialization}
Similarly, we set \(\text{fan\_in}=n_{\mathrm{params}}\):
\begin{equation}
    w_i \sim \mathcal{N}\!\Bigl(0,\; \gamma^2 \cdot \tfrac{1}{n_{\mathrm{params}}}\Bigr)
    \label{lecun1}
\end{equation}
\begin{equation}
    w_i \sim \mathcal{U}\Bigl(-\,\gamma\sqrt{\tfrac{3}{n_{\mathrm{params}}}},\;
    \gamma\sqrt{\tfrac{3}{n_{\mathrm{params}}}}\Bigr),
    \label{lecun2}
\end{equation}
where $n_{\mathrm{params}}$ is the whole parameters in the circuits which is $300$ in this paper discussed in Sec. \ref{subsec:heisenberg}.
Because this variance is smaller than He initialization, it may avoid overly large rotations in early training, which can be beneficial if certain gates saturate (behaving somewhat like sigmoid/tanh).
\subsubsection{Current Limitations}
\textbf{Sigmoid/Tanh Motivation vs.\ Quantum Gates:} 
LeCun normal/uniform was originally intended for sigmoid or tanh activations, optimizing gradient flow in deep networks. Variational circuits do not strictly rely on sigmoidal activation, so its direct benefit is less obvious.

\textbf{Using \(\text{fan-in} = n_{params}\) is Crude:} 
Not all parameters contribute equally to the measured observable. Treating them uniformly can create suboptimal initial states for certain gates or measurement schemes.

\subsubsection{Future Optimizations}
 \textbf{Hybrid Schemes.} 
Depending on their role, some layers could use LeCun-level variance, while others use He (e.g., local vs.\ entangling rotations).

\textbf{Data-Aware Scaling.} 
If one has partial knowledge of expected output states or measurement distributions, scaling \(\gamma\) accordingly can prevent early saturation or plateau.

\subsection{He Initialization}
\subsubsection{Classical Approach}
He initialization \cite{he2015delving} is typically used for ReLU-based networks, which is $\sqrt{2} $ times of LeCun initialization generally, where weights are sampled as
\begin{equation}
    \text{He Normal:}\quad
    w_i \sim \mathcal{N}\!\Bigl(0,\; \tfrac{2}{\text{fan\_in}} \Bigr),
\end{equation}
\begin{equation}
    \text{He Uniform:}\quad
    w_i \sim \mathcal{U}\!\Bigl(-\,\beta,\;\beta\Bigr),
    \;\; \beta = \sqrt{\tfrac{6}{\text{fan\_in}}},
\end{equation}
where $\text{fan\_in}$ indicates the number of input neurons (or input connections) of this layer.
\subsubsection{Proposed Heuristic Quantum Initialization}
In this paper, we assume \(\text{fan\_in}=n_{\mathrm{params}}\) by default, yielding:
\begin{equation}
    w_i \sim \mathcal{N}\!\Bigl(0,\; \gamma^2 \cdot \tfrac{2}{n_{\mathrm{params}}}\Bigr),
    \label{he1}
\end{equation}
\begin{equation}
    w_i \sim \mathcal{U}\Bigl(-\,\gamma\sqrt{\tfrac{6}{n_{\mathrm{params}}}},\;
                                  \gamma\sqrt{\tfrac{6}{n_{\mathrm{params}}}}\Bigr).
                                  \label{he2}
\end{equation}
Again, \(\gamma\) is a global user-chosen factor. While this does not literally correspond to ReLU activations in a quantum circuit, it provides a higher variance than LeCun or Xavier, potentially helping avoid overly small gradients.
\subsubsection{Current Limitations}

\textbf{Misalignment with Classical Activation Theory:} He initialization was derived for ReLU activations to preserve variance in forward/backward passes. VQCs lack direct analogs to ReLU, so the theoretical underpinnings become weaker in purely quantum contexts.

\textbf{Single Variance for All Gate Types:} We use $\sqrt{2 / n_{params}}$ or similar for the entire parameter set. Multi-qubit gates potentially have different gradient behaviors than single-qubit gates; a uniform variance may fail to balance each block's gradient distribution.

\subsubsection{Future Optimizations}

\textbf{Selective Layer Application.} One might only apply “He-like” initialization to single-qubit rotation gates while using smaller variance for multi-qubit entangling blocks, balancing overall gradient flow.

\textbf{Shot Noise Sensitivity.}  High variance might be helpful for exploration but could amplify hardware shot noise when estimating gradients. Finding a noise-aware sweet spot is an open research direction.

\begin{figure*}[htbp]
    \centering
  
    \begin{subfigure}[b]{0.24\textwidth}
        \centering
        \includegraphics[width=\textwidth]{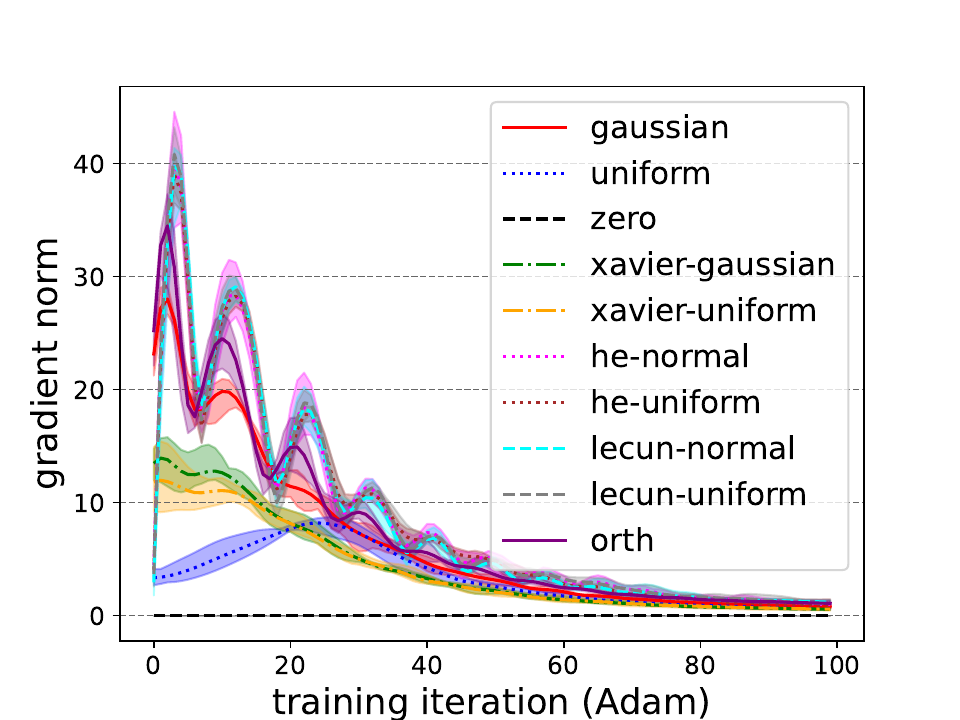}
        \caption{}
        \label{Heisenbergfig:sub1}
    \end{subfigure}
    \hfill
    \begin{subfigure}[b]{0.24\textwidth}
        \centering
        \includegraphics[width=\textwidth]{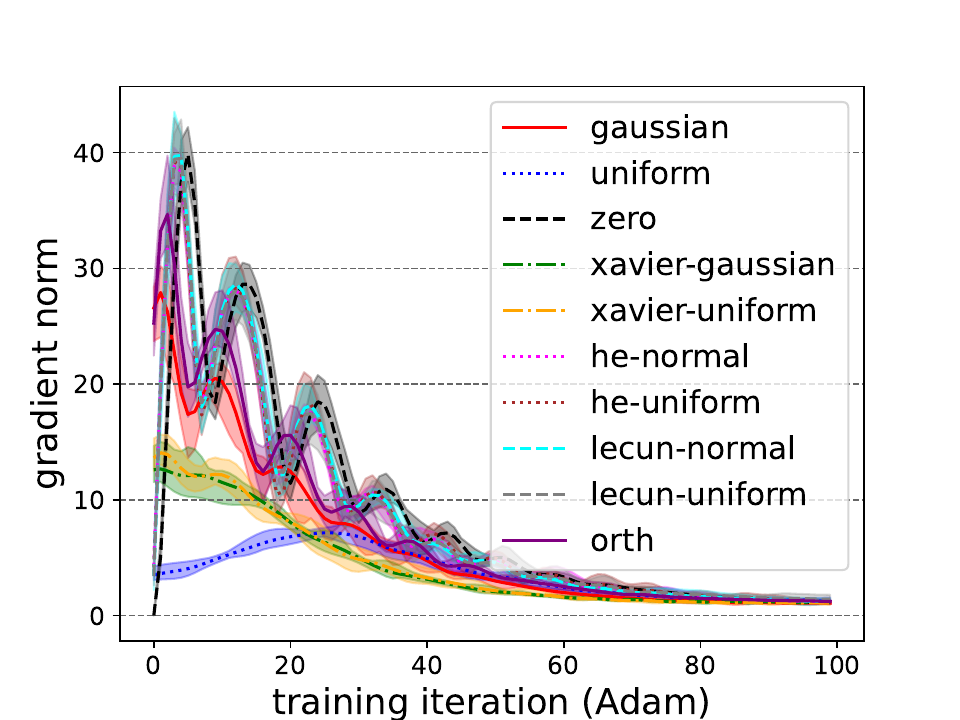}
        \caption{}
        \label{Heisenbergfig:sub2}
    \end{subfigure}
    \hfill
    \begin{subfigure}[b]{0.24\textwidth}
        \centering
        \includegraphics[width=\textwidth]{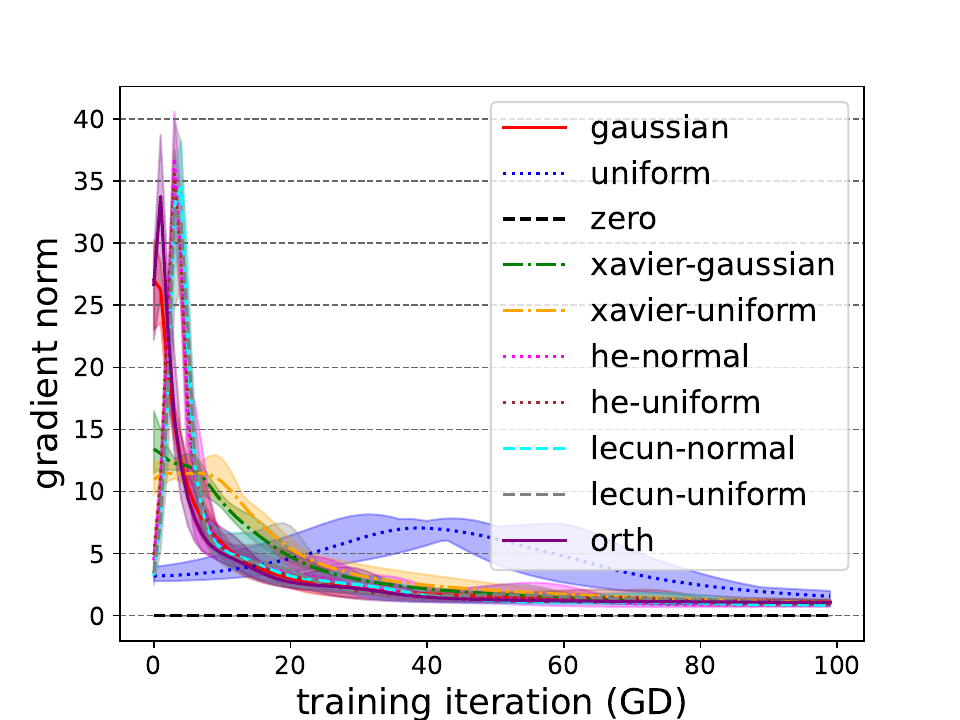}
        \caption{}
        \label{Heisenbergfig:sub3}
    \end{subfigure}
    \hfill
    \begin{subfigure}[b]{0.24\textwidth}
        \centering
        \includegraphics[width=\textwidth]{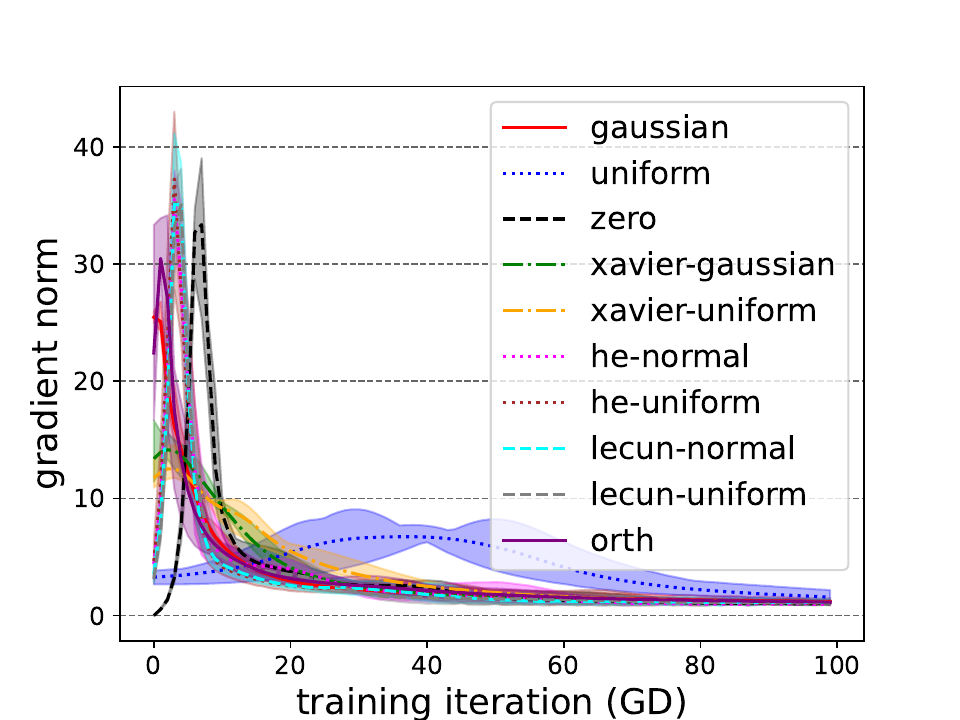}
        \caption{}
        \label{Heisenbergfig:sub4}
    \end{subfigure}

    \vspace{1em}  

    \begin{subfigure}[b]{0.24\textwidth}
        \centering
        \includegraphics[width=\textwidth]{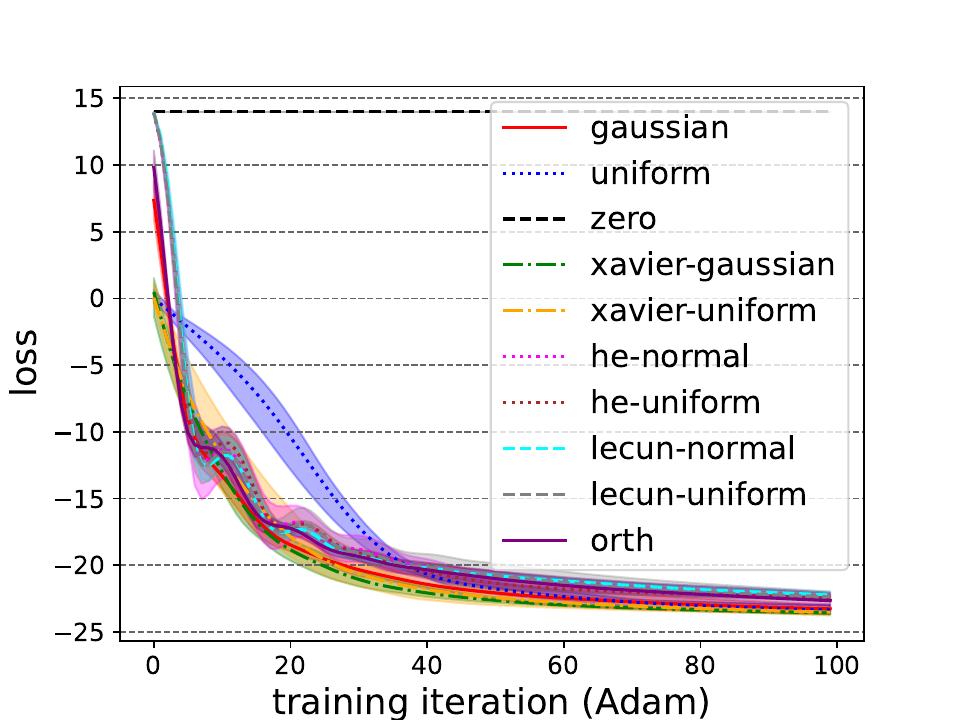}
        \caption{}
        \label{Heisenbergfig:sub5}
    \end{subfigure}
    \hfill
    \begin{subfigure}[b]{0.24\textwidth}
        \centering
        \includegraphics[width=\textwidth]{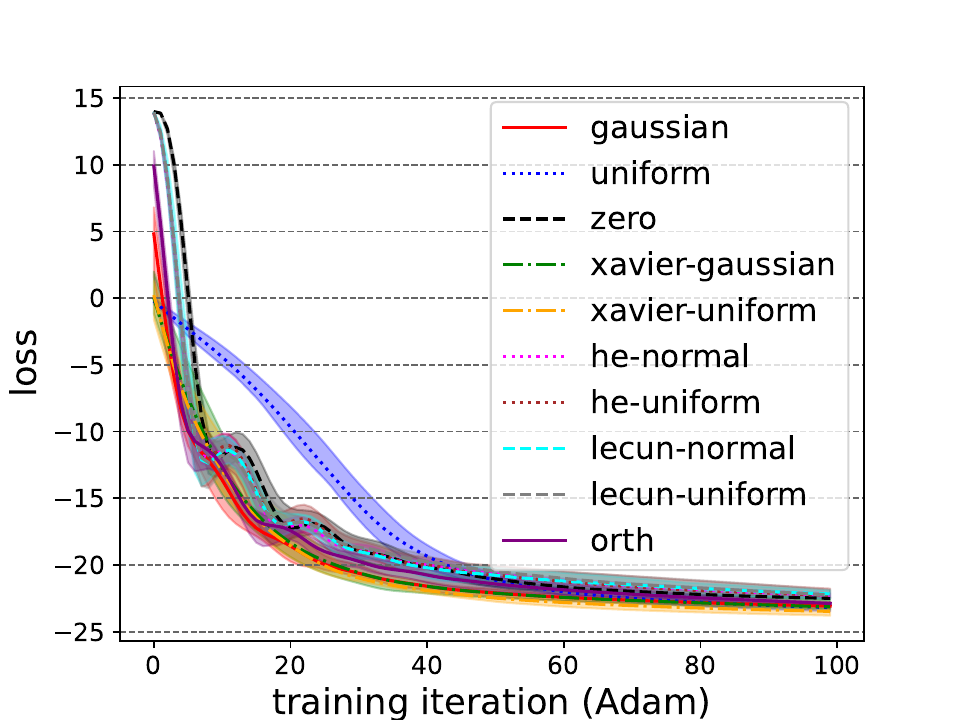}
        \caption{}
        \label{Heisenbergfig:sub6}
    \end{subfigure}
    \hfill
    \begin{subfigure}[b]{0.24\textwidth}
        \centering
        \includegraphics[width=\textwidth]{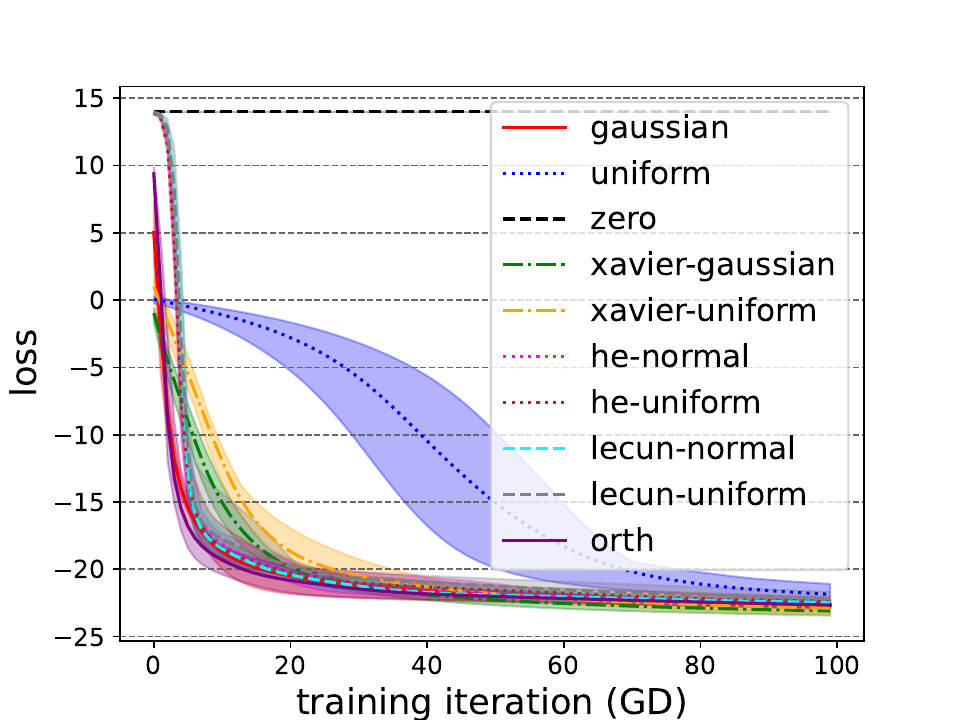}
        \caption{}
        \label{Heisenbergfig:sub7}
    \end{subfigure}
    \hfill
    \begin{subfigure}[b]{0.24\textwidth}
        \centering
        \includegraphics[width=\textwidth]{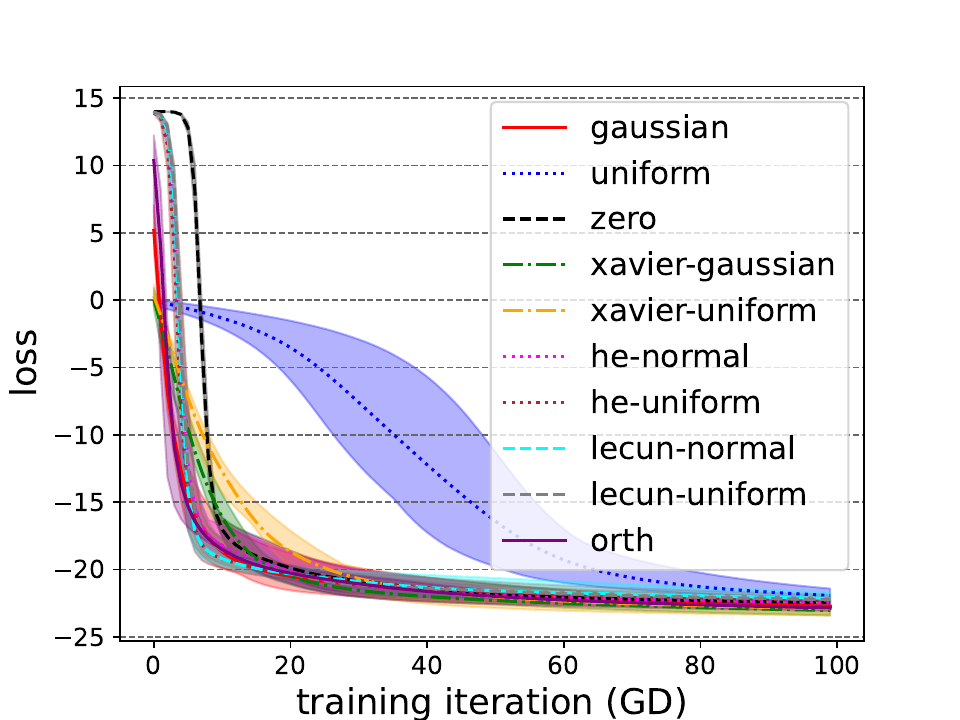}
        \caption{}
        \label{Heisenbergfig:sub8}
    \end{subfigure}

    \caption{Numerical results of finding the ground energy of the Heisenberg model. The first row
shows training results of $l_2$ norm with the Gradient Descent and Adam optimizer, where Figures \ref{Heisenbergfig:sub1} and \ref{Heisenbergfig:sub3} illustrate $l_2$ norm with accurate gradients. Figures \ref{Heisenbergfig:sub2} and \ref{Heisenbergfig:sub4}) show $l_2$ norm with noisy gradients. The second row shows the loss function as Eq. (\ref{eq:HeisenbergHamiltonian}) with the Gradient Descent and Adam optimizer, where Figures \ref{Heisenbergfig:sub5} and \ref{Heisenbergfig:sub7} illustrate loss with accurate gradients. Figures \ref{Heisenbergfig:sub6} and \ref{Heisenbergfig:sub8}) show loss with noisy gradients. Each line denotes the average of $5$ rounds of optimizations.}
    \label{fig:all_five}
    \vspace{-2mm}
\end{figure*}
\subsection{Orthogonal Initialization}
\subsubsection{Classical Approach}
Orthogonal initialization samples a random matrix \(A \in \mathbb{R}^{m \times m}\) from a normal distribution and performs a QR decomposition \((A = QR)\). One then uses rows (or columns) of \(Q\) as weight vectors, ensuring orthogonality.

\subsubsection{Proposed Heuristic Quantum Initialization}
We implement this idea via:  
Generate a random matrix \(A \in \mathbb{R}^{n_{\mathrm{params}} \times n_{\mathrm{params}}}\) from \(\mathcal{N}(0,1)\).
   Factor \(A = QR\) (\(\,Q^\top Q = I\)).
Take the first \(\,n_{\mathrm{times}}\) rows of \(Q\) as mutually orthonormal parameter vectors, then scale by \(\gamma \times \lambda \) if desired where $\lambda$ is the orthonormal rate. This method can be used to provide multiple distinct orthonormal starting points for repeated training runs or for different circuit modules.
\subsubsection{Current Limitations}
\textbf{Orthogonality in Parameter Space vs.\ State Space:} Orthogonal initialization in classical networks helps preserve activation norms across layers. Here, orthogonal vectors in \(\mathbb{R}^{\texttt{n\_params}}\) do not necessarily translate to orthogonal quantum states or beneficial coverage of Hilbert space.

\textbf{Constraint \(\texttt{n\_time} \leq \texttt{n\_params}\):} This paper forms orthogonal vectors by slicing the first \(\texttt{n\_time}\) rows of a QR decomposition. If \(\texttt{n\_time} > \texttt{n\_params}\), no direct orthogonality can be guaranteed for multiple trials. Furthermore, each trial only benefits from orthogonality relative to other trials, not necessarily improving a single trial's gradient behavior.

\subsubsection{Future Optimizations}

\textbf{Complex/Unitary Extensions.} Sampling a complex \(A \in \mathbb{C}^{m \times m}\) and computing a unitary \(U\) might align even more closely with quantum logic, though it is more computationally expensive.

\textbf{Blockwise Orthogonalization.} For large circuits, one could break parameters into blocks and apply orthogonal initialization per block. This might reflect partial entanglement or layered structures.

\textbf{Adaptive Over-Parameterization.} If \(\,n_{\mathrm{times}} > n_{\mathrm{params}}\), partial QR or other factorizations might be necessary to generate sufficiently many distinct vectors without losing orthogonality guarantees.

In conclusion, we have shown how Xavier, He, LeCun, and Orthogonal initialization strategies can be adapted into quantum heuristics. We have completed a preliminary exploration of the heuristics of classical initialization for quantum problems by translating \(\text{fan\_in}\) and \(\text{fan\_out}\) into qubit-based or parameter-based definitions (e.g., setting \(\text{fan\_in}=n_{\mathrm{params}}\) or chunking by qubits).

\begin{figure*}
    \centering
  
    \begin{subfigure}[b]{0.3\textwidth}
        \centering
        \includegraphics[width=\textwidth]{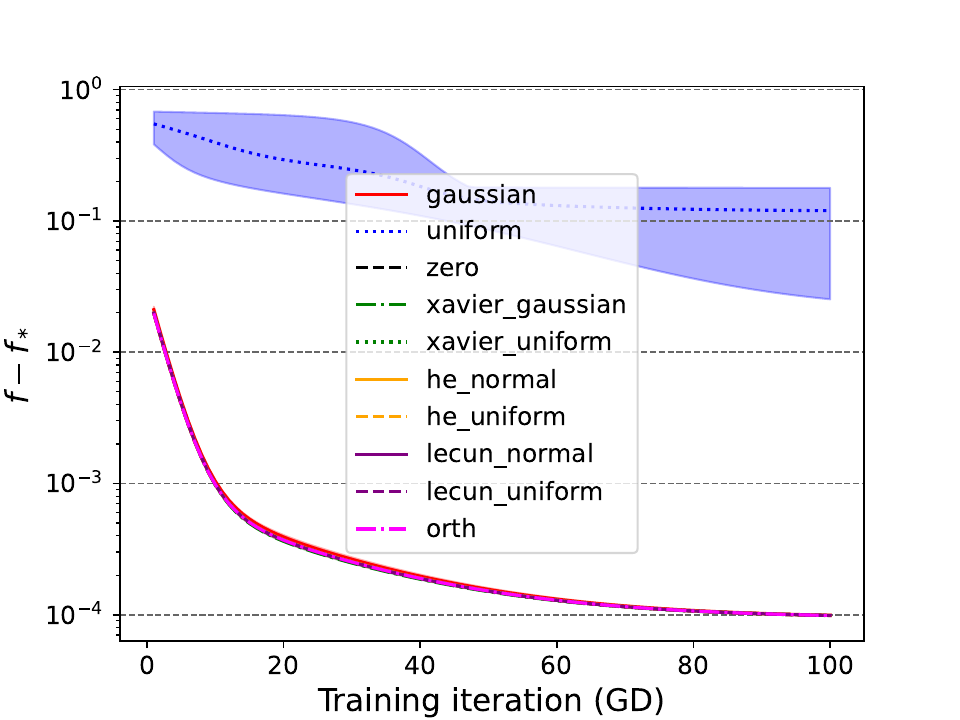}
        \caption{}
        \label{chemenergyfig:sub1}
    \end{subfigure}
    \hfill
    \begin{subfigure}[b]{0.3\textwidth}
        \centering
        \includegraphics[width=\textwidth]{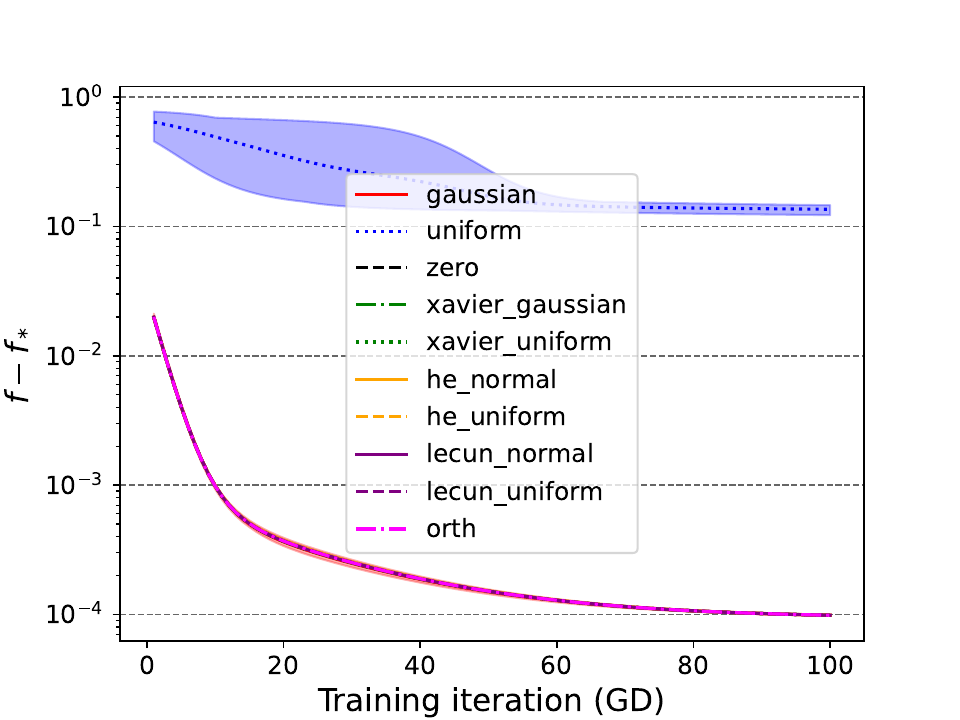}
        \caption{}
        \label{chemenergyfig:sub2}
    \end{subfigure}
    \hfill
    \begin{subfigure}[b]{0.3\textwidth}
        \centering
        \includegraphics[width=\textwidth]{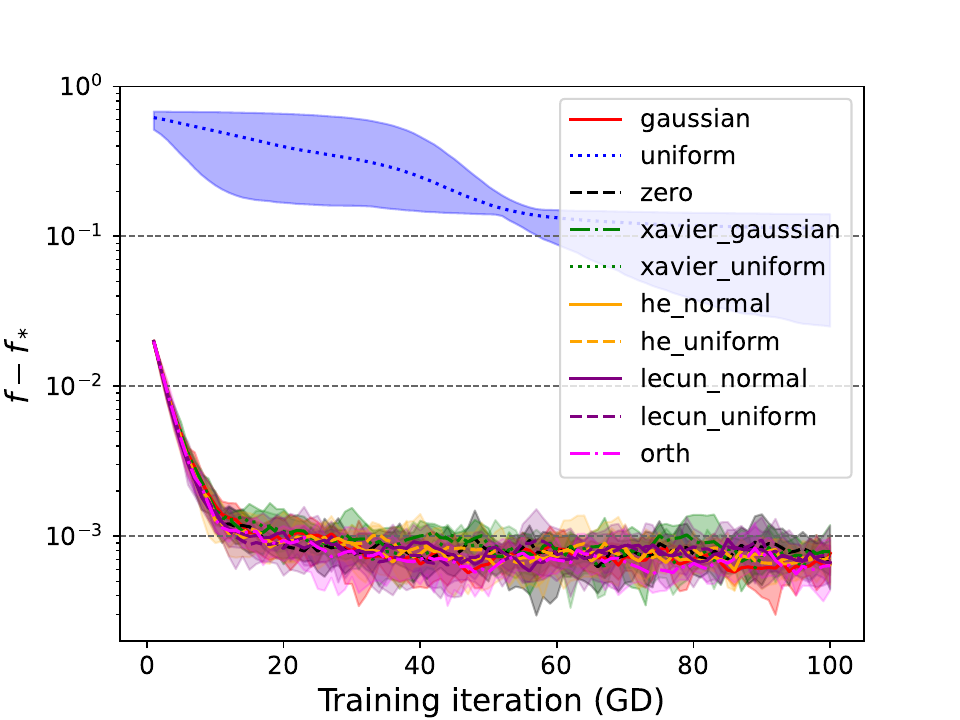}
        \caption{}
        \label{chemenergyfig:sub3}
    \end{subfigure}

    \vspace{1em}  

    \begin{subfigure}[b]{0.3\textwidth}
        \centering
        \includegraphics[width=\textwidth]{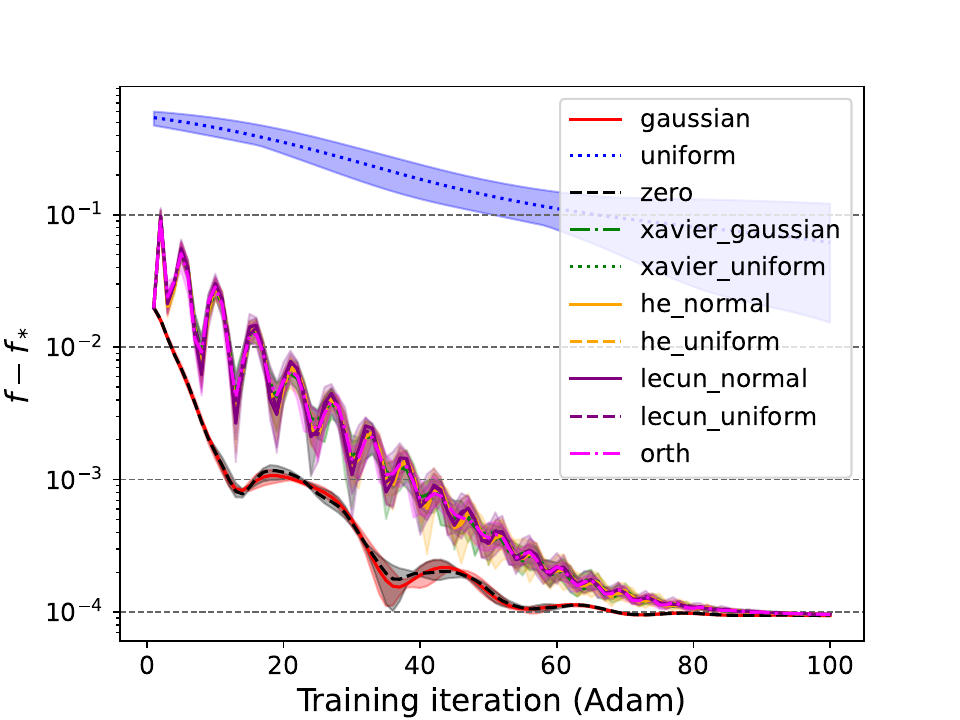}
        \caption{}
        \label{chemenergyfig:sub4}
    \end{subfigure}
    \hfill
    \begin{subfigure}[b]{0.3\textwidth}
        \centering
        \includegraphics[width=\textwidth]{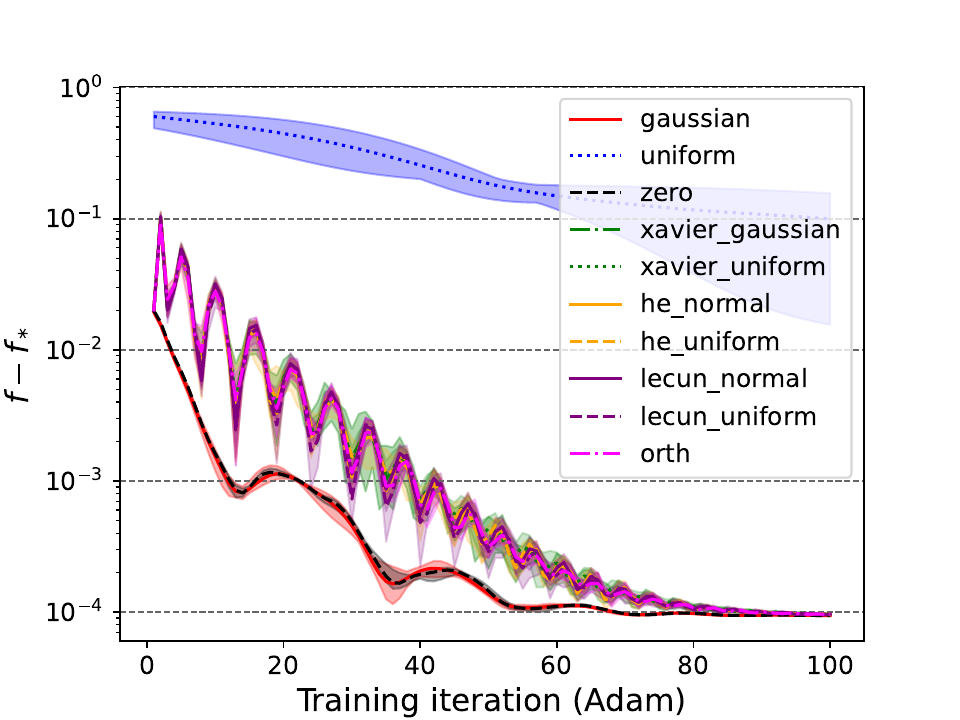}
        \caption{}
        \label{chemenergyfig:sub5}
    \end{subfigure}
    \hfill
    \begin{subfigure}[b]{0.3\textwidth}
        \centering
        \includegraphics[width=\textwidth]{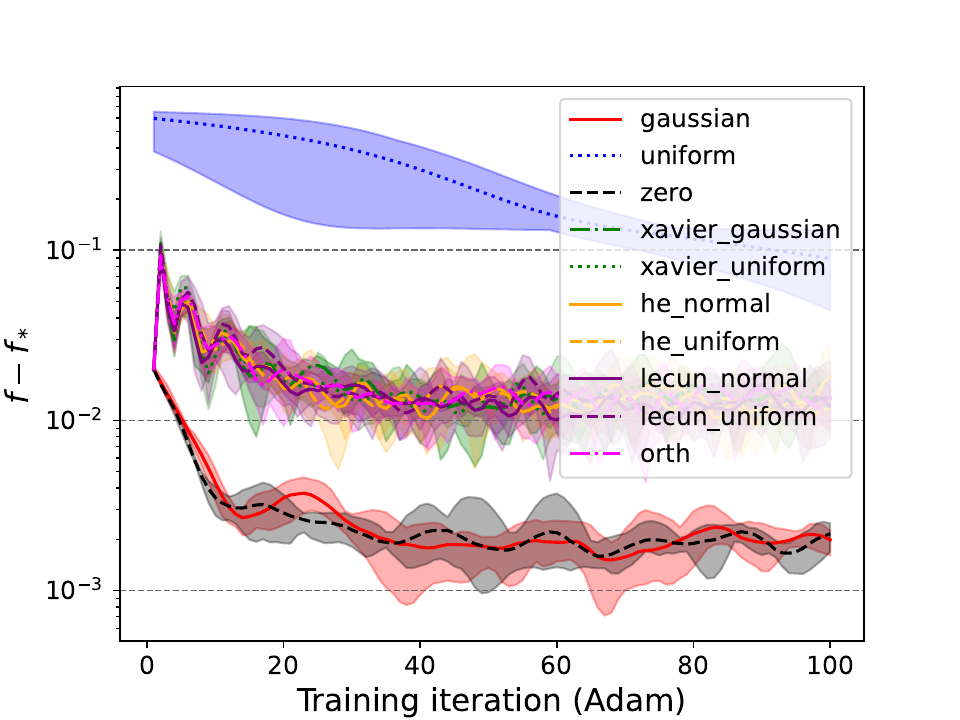}
        \caption{}
        \label{chemenergyfig:sub6}
    \end{subfigure}
\caption{Numerical results of loss $f$ minus global minimum $f_\ast$ of the molecule LiH. The first and second rows show training results with the gradient descent and the Adam optimizer, respectively. The left, the middle, and the right columns show results using accurate gradients, noisy gradients with adaptive-distributed noises, and noisy gradients with constant-distributed noises. The variance of noises in the middle line (Figures \ref{chemenergyfig:sub2} and \ref{chemenergyfig:sub5}) follows Eq. (\ref{eq12}), while the variance of noises in the right line (Figures \ref{chemenergyfig:sub3} and \ref{chemenergyfig:sub6}) is $0.001$. Each line denotes the average of 5 rounds of optimizations.}
    \label{fig:all_chem}
    \vspace{-2mm}
\end{figure*}
\section{Experiments}
\label{sec:experiments}
In this section, we establish the framework of general VQAs and introduce the associated notation.
A typical VQA can be formulated as an optimization problem involving the function
\begin{equation}
\label{eq:generalVQAf}
f(\boldsymbol{\theta}) = \mathrm{Tr}\!\Bigl[
\,O \, V(\boldsymbol{\theta}) \, \rho_{\mathrm{in}} \, V(\boldsymbol{\theta})^\dagger
\Bigr],
\end{equation}
where \(V(\boldsymbol{\theta})\) is a parameterized unitary, \(\rho_{\mathrm{in}}\) is the input quantum state, and \(O\) is the observable. 
The value of \(f(\boldsymbol{\theta})\) depends on both the parameters \(\boldsymbol{\theta}\) and the choice of \(\rho_{\mathrm{in}}\).

\subsection{Heisenberg Model}
\label{subsec:heisenberg}

In this section, we target the ground state and ground energy of the Heisenberg model~\cite{bonechi1992heisenberg}, governed by the Hamiltonian
\begin{equation}
\label{eq:HeisenbergHamiltonian}
H \;=\; \sum_{i=1}^{N-1}\Bigl(X_i X_{i+1} + Y_i Y_{i+1} + Z_i Z_{i+1}\Bigr),
\end{equation}
where \(N\) is the number of qubits.
The operators \(X_i\), \(Y_i\), and \(Z_i\) each act non-trivially on the \(i\)-th qubit (as Pauli \(X\), \(Y\), or \(Z\)) and trivially on all other qubits, \emph{i.e.},
$
X_i = I^{\otimes (i-1)} \otimes X \otimes I^{\otimes (N-i)},
\quad
Y_i = I^{\otimes (i-1)} \otimes Y \otimes I^{\otimes (N-i)},
\quad
Z_i = I^{\otimes (i-1)} \otimes Z \otimes I^{\otimes (N-i)}.
$

We employ the loss function from Eq.~\eqref{eq:generalVQAf} with the input state \(\lvert 0\rangle^{\otimes N}\) and the observable \(H\). 
Minimizing this loss yields the ground energy of \(H\). Our ansatz consists of \(N=15\) qubits arranged in \(L = 10\) layers of \(\mathrm{R_y}\,\mathrm{R_x}\,\mathrm{CZ}\) blocks. 
In each block, we first apply \(\mathrm{CZ}\) gates to all neighboring qubit pairs \(\{(1,2),\dots,(N,1)\}\), followed by \(\mathrm{R_x}\) and \(\mathrm{R_y}\) rotations on each qubit.
This circuit has \(300\) tunable parameters in total.

Since each term of \(H\) contains at most \(S=2\) non-identity Pauli operators, the setup corresponds to \((S,K)=(2,18)\). 
Accordingly, $\gamma^2$ is set in \cite{zhang2022escaping} as 
$
\gamma^2 \;=\; \frac{1}{4\,S\,(K+2)} \;=\; \frac{1}{160}.
$

In this experiment, we train VQAs with gradient descent (GD) and Adam optimizers.
Learning rates are set to $0.01$ for GD and Adam cases, as shown in Figure. \ref{fig:all_five}. The figure shows that the proposed heuristic Xavier initialization shows better convergence utilizing the Adam optimizer and reduced variance over the course of training, reflecting enhanced robustness and reliability in convergence. Meanwhile, the Gaussian initialization is defined as follows \cite{zhang2022escaping}: 
\begin{equation}
    w_i \sim \mathcal{N}\!\Bigl(0,\; \gamma^2\Bigr).
    \label{gaussian1}
\end{equation}
So He and LeCun initialization defined in Eq. (\ref{lecun1}, \ref{lecun2}) and Eq. (\ref{he1}, \ref{he2}) is quite similar to Eq. (\ref{gaussian1}) when $n_{params}$ is $300$. Therefore, future research should aim to define unique $n_{params}$ in deep VQCs.

\begin{figure*}
    \centering
  
    \begin{subfigure}[b]{0.3\textwidth}
        \centering
        \includegraphics[width=\textwidth]{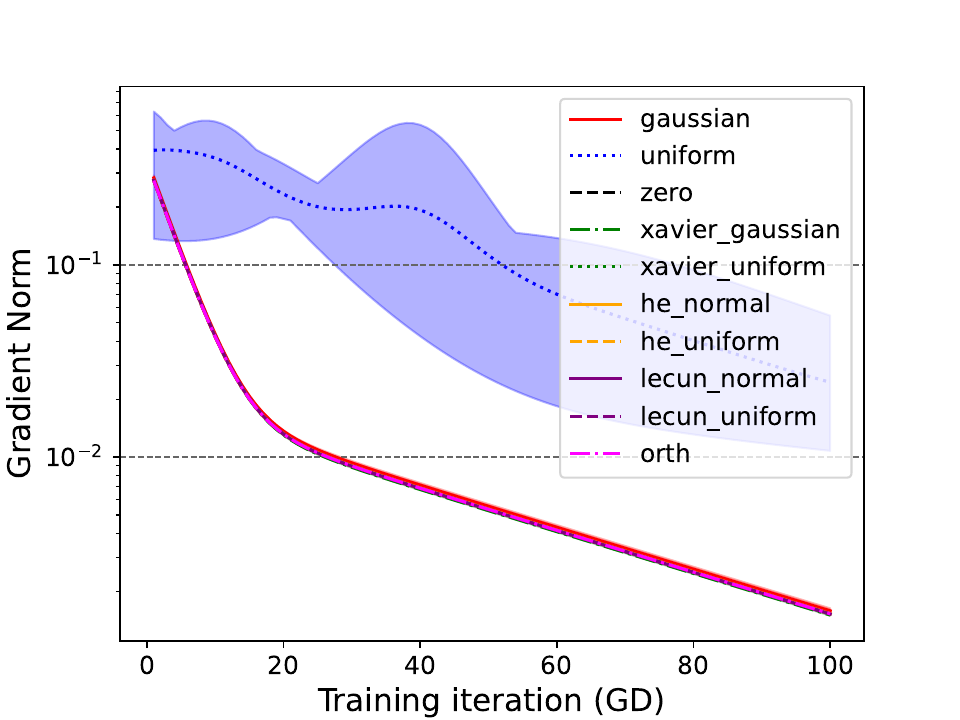}
        \caption{}
        \label{chemnormfig:sub1}
    \end{subfigure}
    \hfill
    \begin{subfigure}[b]{0.3\textwidth}
        \centering
        \includegraphics[width=\textwidth]{Figureschem/chem_gradnorm_gd_10_24_0_0.pdf}
        \caption{}
        \label{chemnormfig:sub2}
    \end{subfigure}
    \hfill
    \begin{subfigure}[b]{0.3\textwidth}
        \centering
        \includegraphics[width=\textwidth]{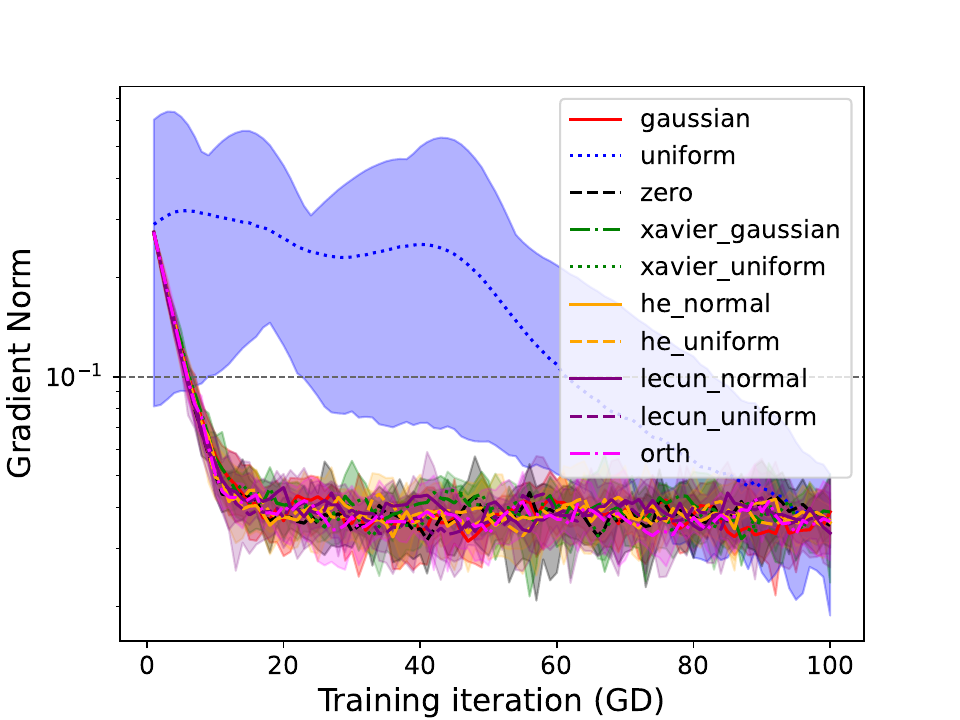}
        \caption{}
        \label{chemnormfig:sub3}
    \end{subfigure}

    \vspace{1em}  

    \begin{subfigure}[b]{0.3\textwidth}
        \centering
        \includegraphics[width=\textwidth]{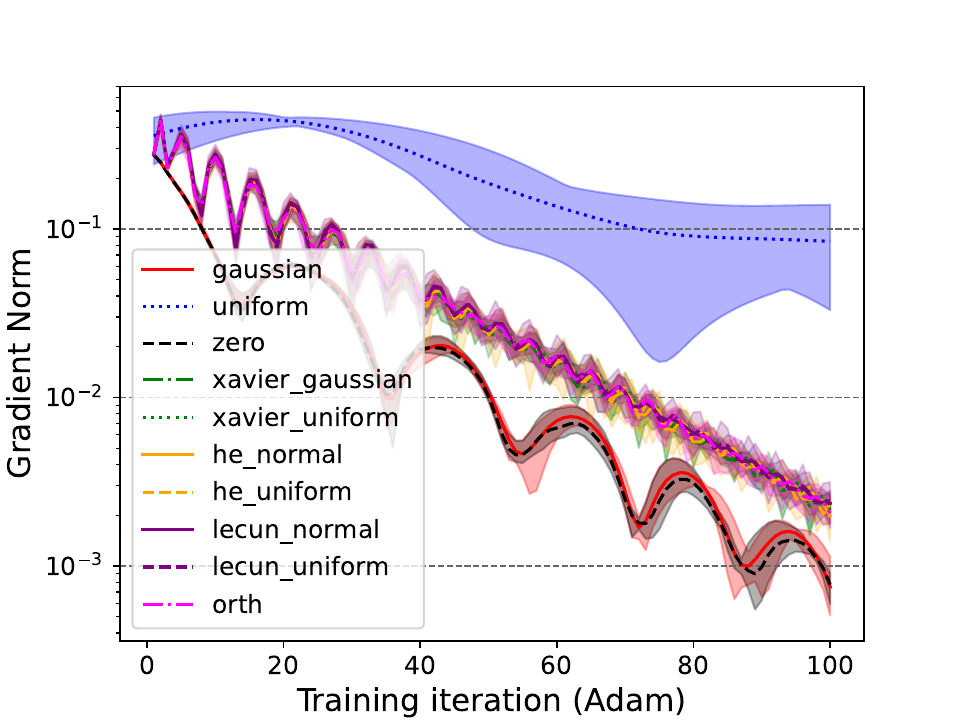}
        \caption{}
        \label{chemnormfig:sub4}
    \end{subfigure}
    \hfill
    \begin{subfigure}[b]{0.3\textwidth}
        \centering
        \includegraphics[width=\textwidth]{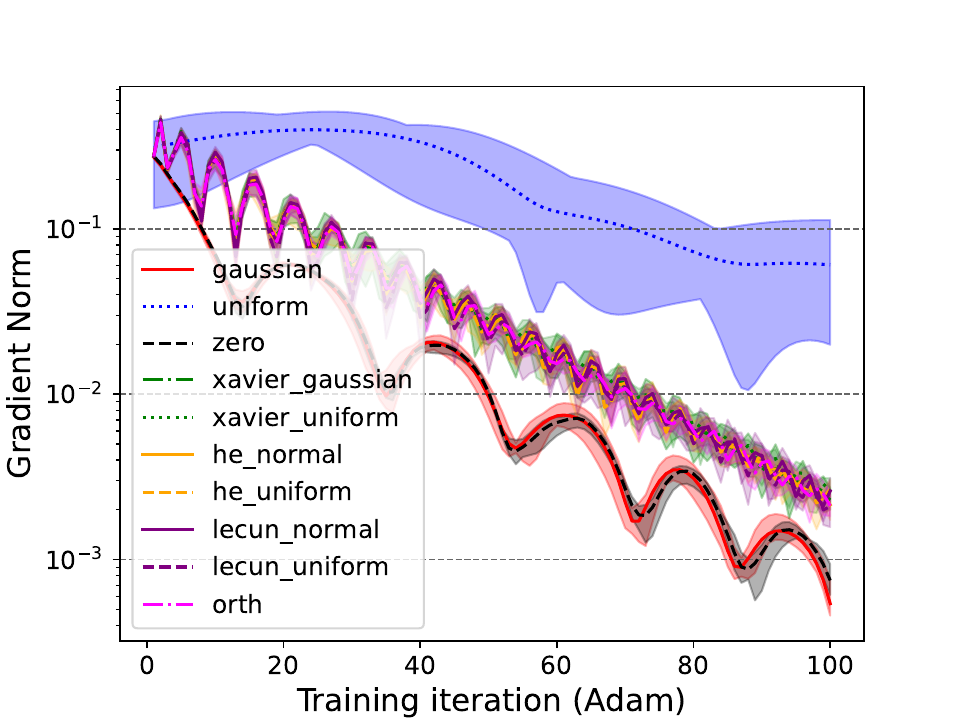}
        \caption{}
        \label{chemnormfig:sub5}
    \end{subfigure}
    \hfill
    \begin{subfigure}[b]{0.3\textwidth}
        \centering
        \includegraphics[width=\textwidth]{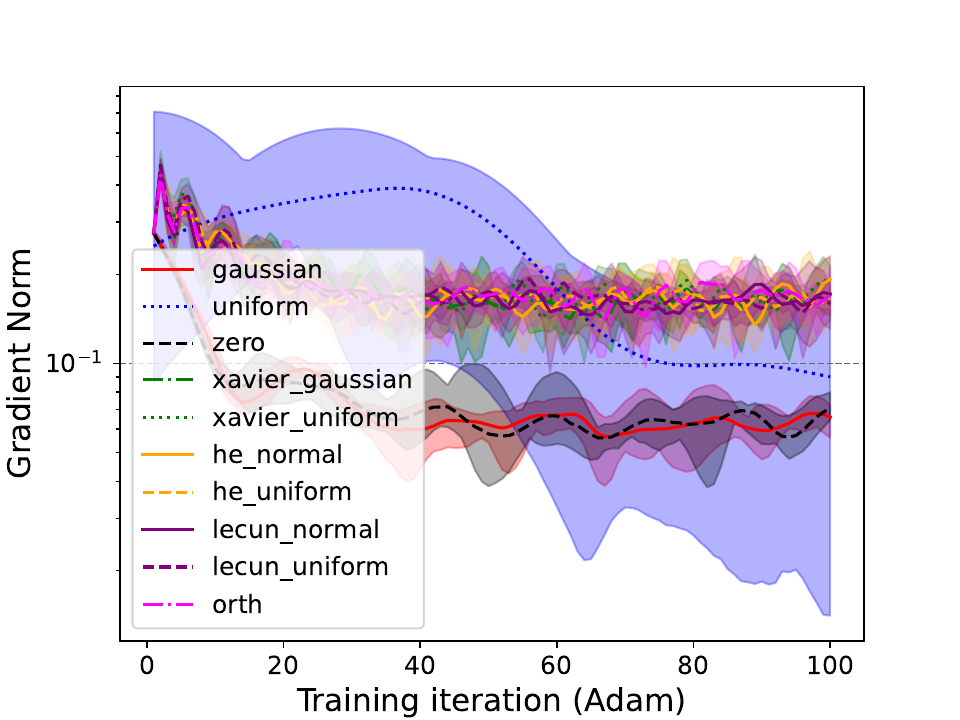}
        \caption{}
        \label{chemnormfig:sub6}
    \end{subfigure}

    \caption{Numerical results of finding the ground energy of the molecule LiH. The first and second rows show training results with the gradient descent and the Adam optimizer, respectively. The left, the middle, and the right columns show results using accurate gradients, noisy gradients with
adaptive-distributed noises, and noisy gradients with constant-distributed noises. The variance of noises in the middle line (Figures \ref{chemnormfig:sub2} and \ref{chemnormfig:sub5}) follows Eq. (\ref{eq12}), while the variance of noises in the right line (Figures \ref{chemnormfig:sub3} and \ref{chemnormfig:sub6}) is $0.001$. Each line denotes the average of 5 rounds of optimizations.}
    \label{fig:all_chemnorm}
    \vspace{-2mm}
\end{figure*}

\subsection{LiH molecule} In this second task, we target the ground state and ground energy of the Lithium Hydride (LiH) molecule by employing a variational quantum algorithm with a Hartree--Fock (HF) reference state and Givens rotation gates following settings in \cite{zhang2022escaping}. Specifically, we set the number of active electrons to \(n_e = 2\) and the number of free spin orbitals to \(n_0 = 10\), yielding \(N = n_0\) qubits. The HF reference state is 
\[
  \ket{\phi_{\mathrm{HF}}} 
  \;=\; \bigl(\ket{1}\bigr)^{\otimes n_e}
         \;\otimes\; 
         \bigl(\ket{0}\bigr)^{\otimes (n_0 - n_e)},
\]
and we construct the parameterized circuit 
$
  V_{\mathrm{Givens}}(\boldsymbol{\theta}) 
  \;=\; 
  \prod_{i=1}^{24} 
  R_{\mathrm{Givens}}^{(i)}(\theta_i)
$
from \(L = 24\) distinct Givens rotation gates, each acting on either \(2\) or \(4\) qubits.

Following \cite{delgado2021variational}, we denote the LiH Hamiltonian by \(H_{\mathrm{LiH}}\). The corresponding loss function,
\begin{equation}
\label{eq11}
f(\boldsymbol{\theta})
\;=\;
\mathrm{Tr}\!\Bigl[
  H_{\mathrm{LiH}}\,
  V_{\mathrm{Givens}}(\boldsymbol{\theta})\,
  \ket{\phi_{\mathrm{HF}}}\bra{\phi_{\mathrm{HF}}}\,
  V_{\mathrm{Givens}}(\boldsymbol{\theta})^\dagger
\Bigr],
\end{equation}
is minimized to obtain LiH's ground energy.

Because each Givens rotation on two qubits is labeled \(\bigl(h,a\bigr) = (2,2)\) and on four qubits is \(\bigl(h,a\bigr) = (8,8)\), we set the variance following \cite{zhang2022escaping} to
$\gamma^2 
  \;=\; 
  \frac{8^3 \times \tfrac{1}{2}}{48 \times 8^2 \times 24},
$
matching the \(\bigl(L,h,a,e\bigr) = (24,8,8,\tfrac12)\) case.
Following the approach in the Heisenberg model, we consider both accurate gradients and noisy gradients with a fixed noise variance of \(0.001\).
Additionally, we analyze a noisy scenario with adaptive noise, where the variance of noise on each partial derivative at iteration \(t\) following \cite{zhang2022escaping} is
\begin{equation}
\label{eq12}
\gamma^2
\;=\;
\frac{1}{96 \,\times\, 24 \,\times\, 8^2}
\;\|\!H_{\mathrm{LiH}}\|\!^2 
\;\left(\frac{\partial f}{\partial \theta}\right)^{\!2}
\Bigg|_{\theta=\theta^{(t)}},
\end{equation}
so that the noise level adapts to the magnitude of the current gradient. Learning rates are set to $0.1$ and $0.01$ for GD and Adam cases, respectively. In this case, $n_{param}$ in the proposed heuristic quantum Xavier, He, and LeCun initialization is defined as the excitations in the molecule LiH (24).

However, one should note that this mapping is purely formal and does not have a direct physical meaning. In quantum chemistry, ``excitations'' usually refer to allowed electronic transitions or the dimension of an active space in a many-body configuration. Simply equating this number to $n_{\mathrm{param}}$ used in classical initialization formulas can lead to curves that appear confusing or difficult to interpret physically, as shown in Figures~\ref{fig:all_chem} and~\ref{fig:all_chemnorm}.

Because the notion of ``excitations'' in LiH does not fully capture the depth or connectivity of the underlying parameterized quantum circuit, the correlation between $n_{\mathrm{param}}$ and the observed training performance (gradient norms, convergence rates, etc.) may not be straightforward. 

To make the interpretation of such curves more physically meaningful, one possible direction is to introduce a refined definition of $n_{\mathrm{param}}$ that better reflects the actual degrees of freedom in the variational circuit. Incorporate the structure and depth of the quantum ansatz (number of layers, type of gates) or the entangling strategy among qubits. These considerations would lead to a parameter definition that is more consistent with the quantum architecture and, thus, potentially yield curves whose trends are easier to correlate with chemistry or algorithmic insights.

\section{Conclusion}
\label{sec:conclusions}
While Gaussian, Zero, and Uniform initializations are simple baselines, the additional Xavier, He, LeCun, and Orthogonal strategies offer refined parameter distributions drawn from classical deep-learning experience. However, their effectiveness in quantum circuits hinges on carefully adapting each method to reflect the structure of gates, measurement types, quantum noise, and device characteristics. These heuristics may be refined in future work to better fit genuine quantum norms and mitigate phenomena such as barren plateaus. In particular, future research should incorporate numerical experiments with quantum noise and evaluate more complex experimental scales with deeper quantum circuits, in order to further assess and improve the practical benefits of these methods. The initialization-method code will be made available upon request for further reference and exploration.

\clearpage

\newpage

\bibliographystyle{IEEEtran}
\bibliography{reference}

\end{document}